\documentclass[aps,prd, twocolumn,superscriptaddress,nofootinbib,preprintnumbers]{revtex4-1}
\usepackage{amsmath}
\usepackage{graphicx}
\usepackage{subfigure}
\usepackage{amssymb}
\usepackage{xcolor}
\usepackage{multirow}
\usepackage{cancel}
\usepackage{color}
\usepackage{ulem}
\usepackage{listings}
\usepackage{xcolor}
\usepackage{float}
\usepackage{slashed}

\usepackage{tikz}
\usetikzlibrary{arrows,shapes}
\usetikzlibrary{trees}
\usetikzlibrary{matrix,arrows} 				
\usetikzlibrary{positioning}				
\usetikzlibrary{calc,through}				
\usetikzlibrary{decorations.pathreplacing}  
\usepackage{pgffor}							

\usetikzlibrary{decorations.pathmorphing}	
\usetikzlibrary{decorations.markings}
\tikzset{
    vector/.style={decorate, decoration={snake}, draw},
	provector/.style={decorate, decoration={snake,amplitude=2.5pt}, draw},
	antivector/.style={decorate, decoration={snake,amplitude=-2.5pt}, draw},
    fermion/.style={draw=black, postaction={decorate},
        decoration={markings,mark=at position .55 with {\arrow[draw=black]{>}}}},
    fermionbar/.style={draw=black, postaction={decorate},
        decoration={markings,mark=at position .55 with {\arrow[draw=black]{<}}}},
    fermionnoarrow/.style={draw=black},
    gluon/.style={decorate, draw=black,
        decoration={coil,amplitude=4pt, segment length=5pt}},
    scalar/.style={dashed,draw=black, postaction={decorate},
        decoration={markings,mark=at position .55 with {\arrow[draw=black]{>}}}},
    scalarbar/.style={dashed,draw=black, postaction={decorate},
        decoration={markings,mark=at position .55 with {\arrow[draw=black]{<}}}},
    scalarnoarrow/.style={dashed,draw=black},
    electron/.style={draw=black, postaction={decorate},
        decoration={markings,mark=at position .55 with {\arrow[draw=black]{>}}}},
	bigvector/.style={decorate, decoration={snake,amplitude=4pt}, draw},
}

\tikzstyle{block} = [draw, rectangle, 
    minimum height=3em, minimum width=6em]

\usepackage[colorlinks,citecolor=blue]{hyperref}
\usepackage{amsmath}
\usepackage{wrapfig}

\usepackage[T1]{fontenc} 
\usepackage[utf8]{inputenc} 
\usepackage{times}

\newcommand{\be}{\begin{equation}}
\newcommand{\ee}{\end{equation}}
\newcommand{\beq}{\begin{equation}}
\newcommand{\eeq}{\end{equation}}
\newcommand{\bea}{\begin{eqnarray}}
\newcommand{\eea}{\end{eqnarray}}
\newcommand{\besp}{\begin{equation}\begin{split}}
\newcommand{\eesp}{\end{split}\end{equation}}

\newcommand{\Eq}[1]{Eq.~(\ref{#1})}

\newcommand{\Dfbd}{\mathord{\buildrel{\lower3pt\hbox{$\scriptscriptstyle\leftrightarrow$}}\over {D}_{\mu}}}
\newcommand{\ave}[1]{\left\langle #1\right\rangle}

\def\mL{\mathcal{L}}

\def\mO{\mathcal{O}}

\def\Z{\mathbb{Z}}

\def\0{\textbf{0}}
\def\1{\textbf{1}}
\def\2{\textbf{2}}
\def\3{\textbf{3}}
\def\4{\textbf{4}}
\def\5{\textbf{5}}
\def\6{\textbf{6}}
\def\7{\textbf{7}}
\def\8{\textbf{8}}
\def\9{\textbf{9}}

\def\d{\text{d}}

\begin{document}

\title{A collider test of nano-Hertz gravitational waves from pulsar timing arrays}

\author{Shao-Ping Li}
\email{spli@ihep.ac.cn}
\affiliation{Institute of High Energy Physics, Chinese Academy of Sciences, Beijing 100049, China}

\author{Ke-Pan Xie}
\email{kpxie@buaa.edu.cn}
\affiliation{School of Physics, Beihang University, Beijing 100191, P. R. China}

\begin{abstract}

A cosmic first-order phase transition (FOPT) occurring at MeV-scale provides an attractive explanation for the nano-Hertz gravitational wave (GW) background, which is indicated by the recent pulsar timing array data from the NANOGrav, CPTA, EPTA and PPTA collaborations. We propose this explanation can be further tested at the colliders if the hidden sector couples to the Standard Model sector via the Higgs portal. Through a careful analysis of the thermal history in the hidden sector, we demonstrate that in order to explain the observed GW signal, the portal coupling must be sizable so that it can be probed through Higgs invisible decay at the LHC or future lepton colliders such as CEPC, ILC, and FCC-ee. Our research offers a promising avenue to uncover the physical origin of the nano-Hertz GWs through particle physics experiments.

\end{abstract}

\maketitle

\newpage
\section{Introduction}

The observation of gravitational waves (GW) from the merger of binary black hole~\cite{LIGOScientific:2016aoc} opens the gate of GW astronomy, enabling the exploration of the Universe through messengers apart from electromagnetic waves and neutrinos. Due to the weakness of gravity, GW can carry information from the very early stage of the Universe, opening the window to detect the important processes prior to the Big Bang nucleosynthesis (BBN) and Cosmic Microwave Background (CMB), and probe new physics beyond the Standard Model (SM). Recently, four pulsar timing array (PTA) collaborations, namely NANOGrav~\cite{NANOGrav:2023gor}, CPTA~\cite{Xu:2023wog}, EPTA~\cite{Antoniadis:2023ott} and PPTA~\cite{Reardon:2023gzh}, release their new datasets, showing compelling evidences of a stochastic GW background that peaks at $\mO(10^{-8})$ Hz. While this can be interpreted as GWs coming from inspiraling supermassive black hole binaries (SMBHBs), new physics also provides attractive explanations~\cite{NANOGrav:2023hvm,Antoniadis:2023zhi,EuropeanPulsarTimingArray:2023qbc}. Intriguingly, Bayesian analysis of the data favors new physics models over the conventional SMBHB interpretation~\cite{NANOGrav:2023hvm}.

One promising new physics explanation of the nano-Hertz GW  is a first-order phase transition (FOPT) that happens at the MeV-scale temperature, which has been studied in the literature~\cite{Han:2023olf,Megias:2023kiy,Fujikura:2023lkn,Zu:2023olm,Athron:2023mer,Yang:2023qlf,Addazi:2023jvg} (also see Refs.~\cite{Kobakhidze:2017mru, Arunasalam:2017ajm, Addazi:2020zcj, Bian:2020urb, Nakai:2020oit,Ratzinger:2020koh, Borah:2021ftr, Freese:2022qrl, Ashoorioon:2022raz, Bringmann:2023opz, Madge:2023cak}).\footnote{Other possible explanations  include, e.g.,  models involving domain walls~\cite{Guo:2023hyp,Kitajima:2023cek,Bai:2023cqj,Blasi:2023sej,Sakharov:2021dim}, cosmic strings~\cite{Ellis:2023tsl,Wang:2023len,Kitajima:2023vre,Lazarides:2023ksx}, inflationary or scalar-induced physics~\cite{Vagnozzi:2023lwo,Franciolini:2023pbf,Cai:2023dls,Oikonomou:2023qfz,Wang:2023ost}, black holes~\cite{Yang:2023aak,Ellis:2023dgf,Shen:2023pan,Ghoshal:2023fhh,Broadhurst:2023tus,Inomata:2023zup,Depta:2023qst,Gouttenoire:2023ftk,Huang:2023chx,Gouttenoire:2023nzr}, etc~\cite{Lambiase:2023pxd,Li:2023yaj,Franciolini:2023wjm}.} In this paper, we consider the GWs induced by the FOPT in a minimal perspective, where the hidden sector only contains a scalar and a dark gauge boson, and it couples to the SM sector via Higgs-portal interactions. We pay particular attention to the connection between the GWs and particle physics, emphasizing the importance of collider experiments as an efficient probe to identify the physical origin of the nano-Hertz GWs.

The coupling strength between the SM Higgs and a MeV-scale hidden sector has been severely constrained by the LHC data. Such a weak interaction suggests the hidden sector may decouple from the SM thermal bath at GeV-scale before the FOPT occurs. Following this early decoupling, the hidden sector generally reaches a lower temperature at late times because the number of SM relativistic degrees of freedom (d.o.f.) drops significantly at around 100 MeV due to the QCD confinement, which in turn reheats the SM sector. Typically, a smaller portal coupling leads to an earlier thermal decoupling and a smaller temperature ratio between the hidden and the SM sectors. However, a smaller temperature ratio reduces the energy fraction of the FOPT in the Universe and hence suppresses the GWs. Consequently, in vast parameter space, the GW signal has a positive correlation with the Higgs portal strength; especially, we find that the recent reported stochastic GW background requires a sizable portal interaction that can be probed by the Higgs invisible decay at the LHC and future colliders such as CEPC, ILC and FCC-ee.

In this work, we adopt the minimal model with a gauged dark $U(1)_X$ to demonstrate the connection between collider experiments and GWs. In Section~\ref{sec:model}, we introduce the model and establish the framework for the FOPT calculation. Then in Section~\ref{sec:thermal}, we derive the thermal evolution of the hidden sector and point out the relation to the Higgs invisible decay. Based on these discussions, we perform the numerical analysis for a few benchmark points, and present the results in Section~\ref{sec:results}. Finally, the conclusion is given in Section~\ref{sec:conclusion}.

\section{FOPT and GWs from a minimal dark $U(1)_X$ scenario}\label{sec:model}

We consider a new physics hidden sector that is gauged under a dark $U(1)_X$ symmetry. The relevant Lagrangian is
\be
\mL\supset D_\mu S^\dagger D^\mu S-V(S,H),
\ee
where $H=(G^+,(h+iG^0)/\sqrt{2})$ is the SM Higgs doublet, $S=(\phi+i\eta)/\sqrt{2}$ is the dark scalar field that is a singlet under the SM gauge group but carries unit charge under the $U(1)_X$, and $D_\mu=\partial_\mu-ig_XA'_\mu$ is the dark covariant derivative with $g_X$ being the gauge coupling and $A'_\mu$ the dark gauge boson. The hidden sector is assumed to interact with the SM sector only through the Higgs portal coupling, which is described by the joint potential
\begin{multline}\label{VHS}
V(H,S)=\mu_h^2|H|^2+\mu_s^2|S|^2\\
+\lambda_h|H|^4+\lambda_s|S|^4+\lambda_{hs}|H|^2|S|^2.
\end{multline}
The scalars develop vacuum expectation values (VEVs) at $(h,\phi)=(v_{\rm ew},v_s)$ which break the electroweak (EW) and $U(1)_X$ gauge symmetries. The dark gauge boson then acquires a mass $m_{A'}=g_Xv_s$.

The coefficients $\{\mu_h^2,\mu_s^2,\lambda_h,\lambda_s,\lambda_{hs}\}$ in potential~\Eq{VHS} can be re-parametrized using the physical observables $\{m_h,v_{\rm ew},m_\phi,v_s,\theta\}$, where $\theta$ is the mixing angle between the Higgs boson $h$ and the dark scalar $\phi$. Given $m_h=125.09$ GeV and $v_{\rm ew}=246$ GeV, there are three free parameters in $V(H,S)$, and $|\theta|\ll1$ is required by the Higgs and EW measurements. The details of the re-parametrization of the potential can be found in Appendix~\ref{app:joint}. Here we focus on the MeV-scale hidden FOPT  well below the EW symmetry breaking and hence $h\to v_{\rm ew}$. Therefore, the tree-level potential at zero temperature can be matched to
\be\label{V0}
V_0(\phi)\approx\frac{m_\phi^2}{8 v_s^2}(\phi^2-v_s^2)^2,
\ee
along the $\phi$-direction.

To study the dynamics of the hidden sector in the hot and dense environment of the early Universe, we need to take into account  the finite-temperature corrections. Here we denote the temperatures in the hidden and SM sectors as $T'$ and $T$, respectively, and allow $\xi\equiv T'/T$ to deviate from 1, and $\xi$ itself also varies during the cosmic evolution.\footnote{More concretely, $T$ is the temperature of the SM photon. When $T\approx 2.5$ MeV, the neutrino decouples from the photon thermal bath and eventually develops a different temperature at $T\lesssim0.5$ MeV.} The effective $\phi$-potential at finite temperatures is 
\begin{align}\label{VT}
V_{\rm eff}(\phi,T')=V_0(\phi)+V_{\rm CW}(\phi)+V_T(\phi, T'),
\end{align}
where $V_{\rm CW}\equiv V_1(\phi)+\delta V(\phi)$  is the Coleman-Weinberg potential at zero temperature with the counterterm $\delta V(\phi)$ attached, and $V_T(\phi, T')\equiv V_{1T}(\phi,T')+V_{\rm daisy}(\phi,T')$ is the one-loop thermal correction plus the daisy resummation term. See Appendix~\ref{app:VT} for the complete expressions.\footnote{See Ref.~\cite{Croon:2020cgk} for the theoretical uncertainties of such a perturbative computation of the effective potential.}

FOPT occurs when the Universe transitions from a false vacuum (local minimum) to a true vacuum (global minimum). Initially, the Universe retains at $\phi=0$ due to the behavior of $V_{\rm eff}(\phi,T')\approx(-m_\phi^2+g_X^2T'^2/2)\phi^2/4$ at $\phi\sim0$, which exhibits a valley for high enough $T'$. The $U(1)_X$ symmetry is preserved at this stage. However, as $T'$ decreases, another local minimum at $\phi\neq0$ emerges and becomes the true vacuum. In certain parameter space, a barrier formed mainly by $A'$ in the thermal loop separates the two minima, preventing a smooth transition. In such cases, the Universe undergoes quantum tunneling to the true vacuum, resulting in a $U(1)_X$-breaking FOPT with bubble nucleation and expansion dynamics.

We briefly describe the FOPT and GW calculation framework. Bubbles containing the $U(1)_X$-breaking vacuum start to nucleate at $T'_n$ when the decay probability in a Hubble volume and Hubble time reaches $\mO(1)$, i.e.
\be\label{nucleation}
\Gamma(T'_n)H^{-4}(T_n)\approx1,
\ee
where decay rate per unit volume and unit time is
\be\label{decay_rate}
\Gamma(T')\approx T'^4\left(\frac{S_3}{2\pi T'}\right)^{3/2}e^{-S_3/T'},
\ee
derived by the $T'$-dependent action $S_3$ of the $O(3)$-symmetric bounce solution~\cite{Linde:1981zj}. The Hubble constant is dominated by the SM sector, i.e.
\be\label{H_radiation}
H(T)\approx2\pi\sqrt{\frac{\pi g_*(T)}{45}}\frac{T^2}{M_{\rm Pl}},
\ee
where $M_{\rm Pl}=1.22\times10^{19}$ GeV is the Planck mass and $g_*(T)=g_{*,{\rm SM}}(T)+g'_{*}(T')\xi^4$ includes both the relativistic d.o.f. of the SM and hidden sectors. We have assumed a mild FOPT such that the Hubble constant is dominated by the radiation energy throughout the transition, and hence nucleation usually ensures the completion of the FOPT~\cite{Ellis:2018mja}. Therefore, \Eq{nucleation} can be adopted as the FOPT criterion, and numerically it implies
\begin{equation}\label{FOPT_criterion}
\frac{S_3}{T'_n}\approx4\log\left(\frac{1}{4\pi}\sqrt{\frac{45}{\pi g_*(T_n)}}\frac{M_{\rm Pl}}{T'_{n}}\xi_n^2\right),
\end{equation}
where $T'_n$ is the nucleation temperature in the hidden sector, while $T_n=T'_n/\xi_n$ is the corresponding SM temperature, and $\xi_n\equiv\xi|_{T=T_n}$. For an MeV-scale FOPT, $g_*(T_n)\sim 10$ and $S_3/T'_n\approx 190$.

FOPTs source stochastic GWs via bubble collisions, sound waves and turbulence in the plasma~\cite{Athron:2023xlk}. The key variables describing the GW spectrum are $\{\alpha,\beta/H_n,T_n,v_w\}$, where
\be
\alpha=\frac{1}{g_*(T_n)\pi^2T_n^4/30}\left(T'\frac{\partial\Delta V_{\rm eff}}{\partial T'}-\Delta V_{\rm eff}\right)\Big|_{T'_n},
\ee
with $\Delta V_{\rm eff}<0$ being the energy difference between the true and false vacua. In other words, $\alpha$ is the ratio of the FOPT latent heat to the radiation energy, and it can be factorized as
\be\label{alpha_factor}
\alpha=\frac{g'_*(T'_n)}{g_*(T_n)}\alpha'\times\xi_h^4,
\ee
where $\alpha'$ is the latent heat over the hidden radiation energy, and $g'_*(T'_n)=2+2=4$ is the number of hidden relativistic d.o.f. before the FOPT. By this factorization, we can calculate $\alpha'$ with the hidden sector observables alone and then convert it to $\alpha$ for GW calculation. Normally, $\alpha'\lesssim1$ for a non-supercooled FOPT. On the other hand, the $\beta/H_n$ parameter, defined as the inverse ratio of the FOPT duration $\beta^{-1}$ to the Hubble time $H_n^{-1}\equiv H^{-1}(T_n)$, can be calculated using only the hidden observables
\be
\frac{\beta}{H_n}=T'_n\frac{\d(S_{3}/T')}{\d T'}\Big|_{T'_n},
\ee
due to the cancellation of $\xi_n$ factor in the definition~\cite{Breitbach:2018ddu}.

After nucleation, the bubbles undergo an accelerating expansion driven by the vacuum pressure. In the case of a mild FOPT where $\alpha\lesssim1$, the friction force exerted by  $A'$ and $\phi$ in the plasma quickly balances the vacuum pressure, resulting in a terminal wall velocity $v_w<1$.\footnote{It has been shown that it is very challenging to explain the nano-Hertz GWs with a supercooled FOPT, due to the phase transition completion and reheating issues~\cite{Athron:2023mer}.} As a result, most of the vacuum energy released during the FOPT is transferred to plasma motion. Therefore, GWs are primarily generated by sound waves, with turbulence playing a secondary role, while the contribution from bubble collisions is negligible~\cite{Ellis:2019oqb}. In our research, we adopt $v_w=0.9$ as a benchmark.

The GW spectrum from sound wave at production can be expressed as~\cite{Hindmarsh:2015qta}
\be\label{sw}\begin{split}
\Omega_{{\rm GW},n}=&~\frac{1}{\rho_{c,n}}\frac{\text{d}\rho_{{\rm GW},n}}{\text{d}\log f}\\
\approx&~1.59\times10^{-1}\times v_w\left(\frac{\kappa_{\rm sw}\alpha}{1+\alpha}\right)^2\left(\frac{\beta}{H_n}\right)^{-1}\\
&~\times\left(\frac{f}{f_{\rm sw}}\right)^3\left(\frac{7}{4+3(f/f_{\rm sw})^2}\right)^{7/2},
\end{split}\ee
where $\kappa_{\rm sw}$ is the fraction of vacuum energy that goes to surrounding plasma, which can be derived by resolving the hydrodynamic motion of the fluid~\cite{Espinosa:2010hh}, $\rho_{c,n}=3M_{\rm Pl}^2H_n^2/(8\pi)$ is the total energy density at $T_n$, and the peak frequency is $f_{\rm sw}=2\beta/{\sqrt{3}v_w}$. After the cosmological redshift, the GW spectrum today is~\cite{Breitbach:2018ddu}
\be\begin{split}
\Omega_{\rm GW}(f)=&~\frac{1}{\rho_c}\frac{\text{d}\rho_{\rm GW}}{\text{d}\log f}\\
=&~\left(\frac{a_n}{a_0}\right)^4\left(\frac{H}{H_0}\right)^2\Omega_{{\rm GW},n}\left(\frac{a_0}{a_n}f\right),
\end{split}\ee
where $a_0$ and $a_n$ are the scale factor today and at FOPT, respectively, and $H_0\approx67~{\rm km}/({\rm s}\cdot{\rm Mpc})$ is the current Hubble constant.

Substituting relevant astrophysical constants into the above equations, we obtain the numerical relation
\begin{multline}
\Omega_{\rm GW}(f)h^2\approx1.238\times10^{-5}\times\left(g_{*,{\rm SM}}(T_n)+g'_*(T'_n)\xi_n^4\right)\\
\times\left(\frac{g_{*s,{\rm SM}}(T_0)}{g_{*s,{\rm SM}}(T_n)+g'_{*s}(T'_n)\xi_n^3}\right)^{4/3}\Omega_{{\rm GW},n}\left(\frac{a_0}{a_n}f\right),
\end{multline}
where $h=H_0/[100~{\rm km}/({\rm s}\cdot{\rm Mpc})]$, the $g_*$'s labeled by a subscript ``$s$'' denote the d.o.f. related to entropy, for example $g_{*s,{\rm SM}}(T_0)\approx3.91$ is the entropy d.o.f. today, and $g_{*,{\rm SM}}(T_n)=g_{*s,{\rm SM}}(T_n)=10.75$ for $T_n\gtrsim0.5$ MeV while $g_{*,{\rm SM}}(T_n)\approx3.36$ and $g_{*s,{\rm SM}}(T_n)\approx3.91$ for $T_n\lesssim0.5$ MeV. By this procedure we can calculate the spectrum $\Omega_{\rm GW}(f)h^2$ after solving the thermal history of the hidden sector. The peak frequency after redshift is
\begin{multline}
\left(\frac{a_n}{a_0}\right)f_{\rm sw}\approx1.3\times10^{-8}~{\rm Hz}~\times\\
\frac{1}{v_w}\left(\frac{\beta/H_n}{10}\right) 
\left(\frac{T_n}{10~{\rm MeV}}\right),
\end{multline}
from which we can clearly see the relation between the reported $\sim10^{-8}$ Hz signal and an MeV-scale $T_n$. To include the finite lifetime of the sound wave period, we multiply \Eq{sw} with an extra factor $H \tau_{\rm sw}\leqslant1$~\cite{Guo:2020grp}. The sub-leading turbulence contribution is calculated using numerical results~\cite{Binetruy:2012ze}.

\section{Thermal history of the hidden sector}\label{sec:thermal}

The Higgs-portal interactions mediated by $\lambda_{hs}$ in \Eq{VHS} ensure that the hidden sector remains in equilibrium with the SM sector at high temperatures. In the temperature range ${\rm MeV}\lesssim T\ll  100~{\rm GeV}$, specifically after the EW symmetry breaking and before the FOPT, thermal contact is maintained through the Higgs induced process: $SS^\dagger\to h^*\to$ SM particle pairs. The thermally averaged rate for this process, denoted as $\Gamma_{SS^\dagger}$, scales as $\lambda_{hs}^2T^5/m_h^4$, so that it decreases more rapidly than the Hubble expansion rate $H(T)\propto T^2/M_{\rm Pl}$. When $\Gamma_{SS^\dagger}/H$ drops below 1, the hidden sector decouples from the SM plasma and evolves with its own temperature $T'$, which might differ from the SM sector temperature $T$.

After the decoupling, the hidden sector typically has a lower temperature $T'<T$. This is because the SM sector is reheated when there is a particle species that decouples from the plasma. Such an effect can be calculated via entropy conservation before and after the decoupling, and it is prominent especially at $T\sim100$ MeV when the quarks and gluons are confined into hadrons, and $g_*(T)$ drops drastically from 61.75 to 10.75. In contrast, the hidden sector does not have such reheating effects as it contains only $S$ and $A$. Therefore $\xi_n<1$, which suppresses the $\alpha$ parameter by a factor of $\xi_n^4$, as indicated by \Eq{alpha_factor}; this further suppresses the GW peak signal by
\be
\Omega_{\rm GW}\to \xi_n^8\Omega_{\rm GW},
\ee
which can be inferred from \Eq{sw}. Therefore, a moderate $\xi_n\sim1/2$ can already suppress the GW signal strength by a factor of $10^{-3}$, and hence deriving $\xi_n$ is very important in the calculation of the FOPT GW spectrum.

$\xi_n$ is determined by evolving the hidden and SM sectors below the decoupling temperature $T_{\rm dec}$, which is resolved by
\be\label{Tdec}
\Gamma_{SS^\dagger}(T_{\rm dec})=H(T_{\rm dec}).
\ee
below this temperature, the above equality becomes a less-than sign, and the thermal contact is lost. When resolving $T_{\rm dec}$, we consider the annihilation of $SS^\dagger\to h^*\to$ a pair of SM particles including electrons, muons, pions and kaons, and the relevant meson interaction vertices are taken from Ref.~\cite{Ibe:2021fed}. Below $T_{\rm dec}$, the two sectors evolve separately and the entropy in either sector conserves, and hence
\be\label{xi_h-T}
\xi(T)=\left(\frac{g_{*s,{\rm SM}}(T)}{g_{*s,{\rm SM}}(T_{\rm dec})}\right)^{1/3},
\ee
which is smaller than 1 when $T<T_{\rm dec}$ and $g_{*s}(T)$ drops below $g_{*s}(T_{\rm dec})$.

The zero-temperature scalar-Higgs mixing angle $\theta$ is crucial in evaluating $T_{\rm dec}$, as the relevant annihilation processes are mediated by an off-shell Higgs boson and hence has a $\Gamma_{SS^\dagger}\sim\lambda_{hs}^2T^5/m_h^4$ scaling, while $\lambda_{hs}$ is related to the zero-temperature via
\be
\lambda_{hs}\approx\frac{m_\phi^2-m_h^2}{v_{\rm EW}v_s}\theta.
\ee
See also the details in Appendix~\ref{app:joint}. In general, a larger $\theta$ results in a larger annihilation rate and consequently a $T_{\rm dec}$ closer to $T_n$, which then gives a larger $\xi_n$, relaxing the suppression on the GW signals. Therefore, the observed GW background favors a large $\theta$. However, $\theta$ also controls the Higgs exotic decay partial width by
\be
\Gamma(h\to2\phi)=\frac{\theta^2m_h}{32\pi}\frac{m_h^2}{v_s^2}\left(1+\frac{2m_\phi^2}{m_h^2}\right)^2\sqrt{1-\frac{4m_\phi^2}{m_h^2}},
\ee
and hence is stringently constrained by current experiments.

$\phi$ is not a stable particle in our model, and it decays via the mixing with the Higgs boson. For MeV-scale $\phi$, the decay to $e^+e^-$ final state dominates, and the corresponding lifetime of $\phi$, as we will see in the next section, is $\mO(1)$ s, resulting in Higgs invisible decay signals at colliders. Currently, the ATLAS collaboration sets $\text{BR}(h\to 2\phi)<0.16$ at $95\%$ C.L. by global fits based on the 13 TeV LHC data with an integrated luminosity of 139 fb$^{-1}$~\cite{ATLAS-CONF-2021-053}, and this is interpreted as an upper bound of $|\theta|\lesssim10^{-5}$.  Future detection from the HL-LHC can reach $\text{BR}(h\to 2\phi)\approx 0.019$~\cite{deBlas:2019rxi}, and moreover, the future Higgs factories such as CEPC, ILC or FCC-ee can reach $\text{BR}(h\to2\phi)\sim 10^{-3}$~\cite{Tan:2020ufz,Ishikawa:2019uda,Potter:2022shg,Cerri:2016bew}, increasing the sensitivity of $\theta$ by at least an order of magnitude. Those collider experiments can efficiently probe the parameter space favored by the nano-Hertz GW signals indicated by the PTA data, as we will show in the next section.

\section{Testing the origin of GWs at colliders}\label{sec:results}

\begin{figure*}
	\centering
	\includegraphics[width=0.8\textwidth]{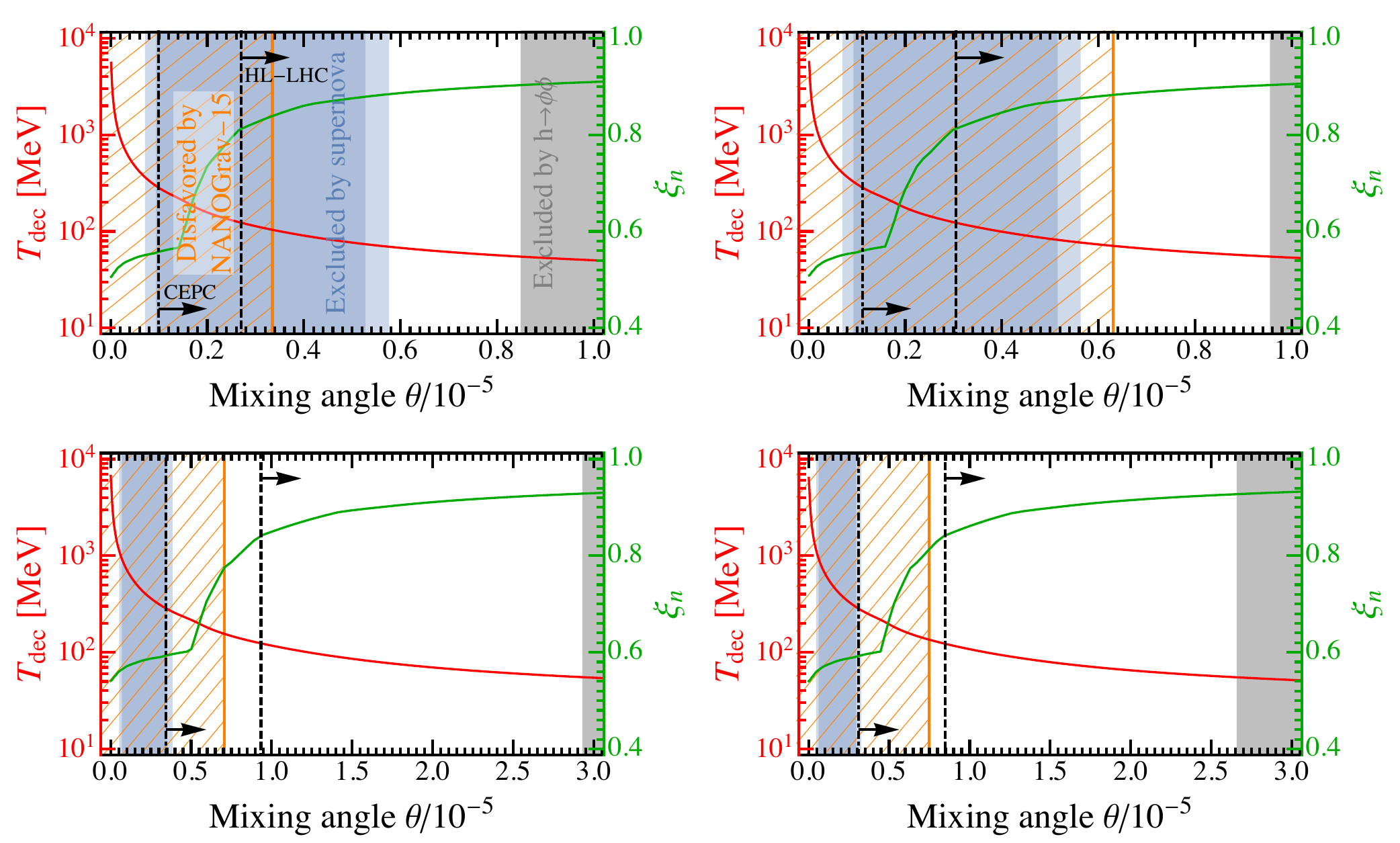}
	\caption{Four BPs: BP1 (top left), BP2 (top right), BP3 (bottom left) and BP4 (bottom right). The $T_{\rm dec}$ and $\xi_n$ as functions of $\theta$ are shown as red and green curves, respectively. The region disfavored by the PTA data is covered by orange mesh, while the region excluded by the LHC data is covered by gray. The supernova-excluded region is covered by blue. The projected reach of HL-LHC and CEPC are shown as gray dashed and dotted-dashed vertical lines, respectively.}\label{fig:Bound_4BP}  
\end{figure*}

\begin{table}\footnotesize
\begin{tabular}{c|c|c|c|c|c|c|c}\hline\hline
 & $m_\phi$ [MeV] & $v_s$ [MeV] & $g_X$ & $\theta_{\rm max}/10^{-5}$  & $ \alpha$  &  $\beta/H_n$  & $T_n$ [MeV] \\ \hline
BP1 & 8.46  & 42.5  & 1.01  & 0.849  &  0.309  & 11.2 & 9.56 \\ \hline
BP2 & 9.16  & 47.9  & 0.981  & 0.955  & 0.269   & 8.17 & 11.2 \\ \hline
BP3 &  23.0 &  147 & 0.892  & 2.93  & 0.523   & 12.3  & 24.3 \\ \hline
BP4 & 31.6  & 133  & 1.13  & 2.66  &  0.684  & 16.8 & 23.9 \\ \hline\hline
\end{tabular}
\caption{The chosen BPs with the $(m_\phi,v_s,g_X)$ values. $\theta_{\rm max}$ is the maximal $\theta$ allowed by the LHC data, and the subsequent $(\alpha,\beta/H_n,T_n)$ is evaluated at $\theta_{\rm max}$.}\label{tab:BPs}
\end{table}

To quantitatively show our results, for a given parameter set of $(m_\phi,v_s,g_X)$, we use the Python package {\tt cosmoTransitions}~\cite{Wainwright:2011kj} to calculate the bounce action $S_3$ and derive $T'_n$ via \Eq{FOPT_criterion}. Then a scan on $\theta$ is performed to determine $T_{\rm dec}$ via \Eq{Tdec} and hence $\xi_n$ via \Eq{xi_h-T}. We choose four benchmark points (BPs) based on the relatively light ($m_\phi\lesssim10$ MeV) and heavy ($m_\phi\gtrsim20$ MeV) scalar scenarios, as listed in Table~\ref{tab:BPs}, and plot $T_{\rm dec}$ and $\xi_n$ as functions of $\theta$ in Fig.~\ref{fig:Bound_4BP}, from which we immediately see the expected negative (positive) correlation of $T_{\rm dec}$ ($\xi_n$) to $\theta$. For the parameter space of interest, $T_{\rm dec}$ is at GeV to MeV scale, while $\xi_n$ varies from 0.5 to 1. We also notice that $\xi_n$  
changes rapidly when $T_{\rm dec}$ crosses $\sim100$ MeV due to the variation of $g_*(T)$ caused by QCD confinement, which can be understood from \Eq{xi_h-T}.

Given $\xi_n$, we can calculate the GW spectrum as described in Section~\ref{sec:model}, mainly based on the discussions around \Eq{sw}. The predicted GW signal curve can then be compared with the PTA data from the NANOGrav~\cite{NANOGrav:2023gor} and CPTA~\cite{Xu:2023wog} collaborations. For the former, we adopt the $f$-$\Omega_{\rm GW}(f)h^2$ data points from Ref.~\cite{NANOGrav:2023hvm}; while for the latter, we use the the best fit point of $f=14$ nHz and transfer ${\rm log}_{10}A=-14.4^{+1.0}_{-2.8}$~\cite{Xu:2023wog} into $\Omega_{\rm GW}h^2$ as an estimate. By varying $\theta$, we see the expected $\xi_n^8$ suppression effect on the GW signal strength. Hence the parameter space favored by the PTA data exhibits a lower limit for $\theta$, as shown in Fig.~\ref{fig:Bound_4BP}, where the disfavored regions are covered by orange mesh. By ``favored by the PTA data'', we require the GW signals to match the first 14 frequency bins of the NANOGrav-15yr data as they are the most accurate and hence have the largest impact on fit quality, following Ref.~\cite{NANOGrav:2023hvm}. Currently, both the SMBHB and FOPT interpretations are consistent with the data, although the latter has a higher Bayesian evidence~\cite{NANOGrav:2023hvm}. Also, the favored FOPT parameter space varies in different fitting setups~\cite{NANOGrav:2023hvm,Ellis:2023oxs,Ghosh:2023aum,Bian:2023dnv}. We expect the future accumulated PTA data could help us to better distinguish the signal origins.

On the other hand, current LHC measurements of Higgs invisible decay set an upper bound $\theta_{\rm max}\sim10^{-5}$ on the mixing angle, and this is shown as the gray shaded regions in Fig.~\ref{fig:Bound_4BP}. In addition, a narrow range of $3.9 \times 10^{-7}\lesssim\theta\lesssim7.0 \times 10^{-6}$ is excluded by the supernova luminosity limit of SN1987A via the nucleon bremsstrahlung process $NN\to NN\phi$~\cite{Dev:2020eam}, as shown in blue shaded regions in the figure. The inner and outer edges correspond to  luminosity limit of $5\times 10^{53}~{\rm erg/s}$ and $3\times 10^{53}~{\rm erg/s}$, respectively.
 
The FOPT temperatures of our BPs are typically larger than a few MeV, and thus are safe under the BBN and CMB constraints~\cite{Bai:2021ibt,Deng:2023seh}. The constraints from ultra-compact minihalo (UCMH) abundance~\cite{Liu:2022lvz,Liu:2023hte} can also be relaxed by adopting a conservative value of the redshift $z_c=1000$ of the last formation of UCHM.  The long-lived light scalar $\phi$ with a life time of $\tau_\phi\sim\mO(1)$ s can affect the BBN process, which in general depends on $m_\phi$, $\tau_\phi$ and the number density $n_\phi$ at decay. Following the analysis of Refs.~\cite{Hufnagel:2018bjp,Depta:2020zbh}, we calculate for each BP the parameter set of $(m_\phi,\tau_\phi,n_\phi)$ and compare with the limits inferred therein. We have checked that all the four BPs are allowed by the BBN constraints.

\begin{figure}[t]
\centering
\includegraphics[scale=0.4]{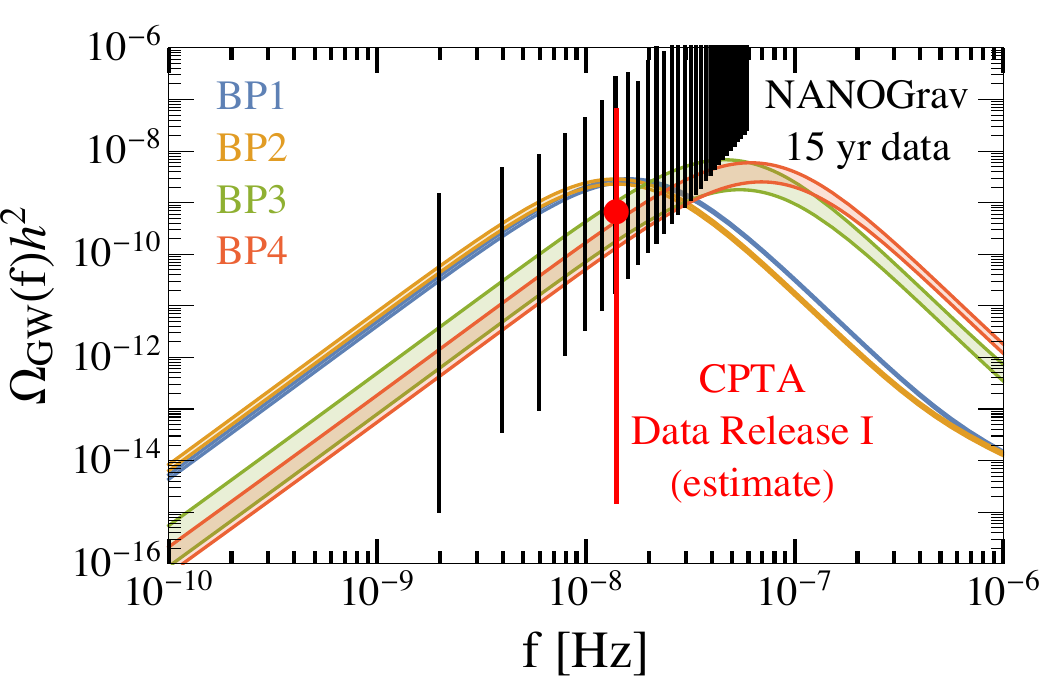}
\caption{Envelopes of the GW spectra for the BPs in Table~\ref{tab:BPs} when varying $\theta$ within the allowed region. The PTA data are explained in the main text.}\label{fig:GW_4BP}  
\end{figure}

Combing the above discussions, the regions that are not covered by neither mesh nor colors are viable for explaining the nano-Hertz GW background while being consistent with existing bounds from particle physics, astrophysics and cosmology, opening a window for future experimental exploration. The envelopes of GW spectra of those four BPs within the allowed $\theta$ regions are plotted in Fig.~\ref{fig:GW_4BP}, and the corresponding parameters $\alpha$, $\beta/H_n$ and $T_n$ for the maximally allowed $\theta_{\rm max}$ are listed in Table~\ref{tab:BPs}.

We represent the projected reach of $\text{BR}(h\to\phi\phi)$ at the HL-LHC and the CEPC as vertical dashed and dotted-dashed black lines, with 1.9\%~\cite{deBlas:2019rxi} and 0.26\%~\cite{Tan:2020ufz} respectively. The HL-LHC has the capability to explore the entire GW-favored parameter space for BP1 and BP2, as well as a portion of BP3 and BP4. The future CEPC can probe entire allowed parameter space for each of the BPs, and anticipated sensitivities of the ILC and the FCC-ee are comparable~\cite{Ishikawa:2019uda,Potter:2022shg,Cerri:2016bew}. Future multi-TeV muon colliders may even reach an invisible decay branching ratio of $\sim10^{-4}$, and hence can also test our scenario~\cite{Ruhdorfer:2023uea}. Therefore, both ongoing and forthcoming collider experiments serve as highly effective means to investigate the physics associated with the nano-Hertz GW background.

\section{Discussion and conclusion}\label{sec:conclusion}
Before conclusion, we would like to emphasize that  the Higgs portal interaction is assumed to be the primary connection between the visible and dark sectors throughout this work. However, there are other possible connections, such as kinetic mixing $\epsilon F'_{\mu\nu}F^{\mu\nu}$ between the dark photon $A'$ and the SM photon $A$. In this case, a value of $\epsilon\gtrsim10^{-8}$ can maintain equilibrium between the two sectors via $SS^\dagger\to A'\to f\bar f$ down to MeV-scale temperatures, with $f$ the SM particles. With $10~\text{MeV}\lesssim m_{A'}\lesssim 100$~MeV favored by PTA data,  the current bounds from colliders, beam dump experiments and astrophysics only  allow for two ranges of the kinetic mixing: $10^{-5}\lesssim\epsilon\lesssim10^{-3}$ and $\epsilon\lesssim10^{-10}$~\cite{Fabbrichesi:2020wbt}, together with some islands $10^{-14}\lesssim\epsilon\lesssim10^{-12}$ and $10^{-17}\lesssim\epsilon\lesssim10^{-16}$ excluded by BBN and CMB observations, respectively~\cite{Fradette:2014sza}. Therefore, the regime $10^{-5}\lesssim\epsilon\lesssim10^{-3}$ keeps the two sectors in thermal equilibrium, and consequently the correlation between Higgs invisible decay and the GW signal strength would be  lost. In this case, alternate probes could be conducted by focusing on the dark photon~\cite{Batell:2022dpx}. Our scenario of Higgs portal dominance applies when the kinetic mixing falls within the $\epsilon\lesssim10^{-10}$ range, or it is forbidden by a $\Z_2$ symmetry with $A'\to-A'$ and $S\to S^\dagger$~\cite{Kanemura:2010sh,Lebedev:2011iq}.

In either case mentioned above, the dark gauge boson $A'$ could be long-lived and contribute to the dark matter relic density at present day. Since $m_{A'}>m_\phi$, the $A'$ will freeze-out via secluded annihilation  $A'A'\to\phi\phi$, with the relic density today scaling like the standard WIMP paradigm:
\be
\Omega_{A'}h^2\approx 0.1\left(\frac{0.1}{g_X}\right)^4 \left(\frac{m_{A'}}{100~{\rm GeV}}\right)^2,
\ee
which is $\sim10^{-12}$ for the parameter space of interest, much smaller than the dark matter relic abundance and hence negligible.

In conclusion, we propose a collider test to the nano-Hertz GW signals reported by the recent PTA experiments. Our idea is based on the extensively studied Higgs-portal model, which consists of a hidden sector that is gauged under a dark $U(1)_X$. We demonstrate that the explanation of the PTA data requires a sizable Higgs portal coupling, which results in a $h\to\phi\phi$ exotic decay that leads to an invisible final state, and can be probed at the colliders. By choosing several BPs as examples, we show that the HL-LHC is able to efficiently probe the parameter space required by the PTA data, while the future lepton colliders such as CEPC can cover almost all the parameter space of interest. Our research shows the   importance of collider experiments in testing the origin of the stochastic GW background.

\section*{Acknowledgements}

We would like to thank Jing Liu and Mengchao Zhang for very useful discussions. S.-P. Li is supported in part by the National Natural Science Foundation of China under grant No. 12141501.

\appendix

\section{The re-parametrization of the joint potential}\label{app:joint}

In the unitary gauge, $H\to h/\sqrt{2}$ and $S\to\phi/\sqrt{2}$, the potential is rewritten as
\be\label{VU}
V(h,\phi)=\frac{\mu_h^2}{2}h^2+\frac{\lambda_h}{4}h^4+\frac{\mu_s^2}{2}\phi^2+\frac{\lambda_s}{4}\phi^4+\frac{\lambda_{hs}}{4}h^2\phi^2,
\ee
which is minimized at the vacuum $(h,\phi)=(v_{\rm ew},v_s)$. Given the mass eigenvalues $m_h$, $m_\phi$ and the mixing angle $\theta$, the coefficients in \Eq{VU} can be expressed as
\begin{multline}\label{reparametrize}
\mu_{h,s}^2=-\frac{1}{4}\Big[m_h^2+m_\phi^2\mp(m_\phi^2-m_h^2)\\
\left(\cos2\theta\mp\left(\frac{v_s}{v_{\rm ew}}\right)^{\pm1}\sin2\theta\right)\Big],
\end{multline}
and
\be\begin{split}
\lambda_h=&~\frac{m_h^2+m_\phi^2-(m_\phi^2-m_h^2)\cos2\theta}{4v_{\rm ew}^2},\\
\lambda_s=&~\frac{m_h^2+m_\phi^2+(m_\phi^2-m_h^2)\cos2\theta}{4v_s^2},\\
\lambda_{hs}=&~\frac{(m_\phi^2-m_h^2)\sin2\theta}{2v_{\rm ew}v_s}.
\end{split}\ee
For small mixing angle $|\theta|\ll1$, the interaction vertices can be read from the following interaction Lagrangian
\begin{multline}
\mL_3\approx-\frac{m_h^2}{2v_{\rm ew}}h^3+\frac{2 m_h^2+m_\phi^2} {2v_{\rm ew}}\theta h^2\phi\\
+\frac{m_h^2 + 2 m_\phi^2}{2v_s}\theta h\phi^2-\frac{m_\phi^2}{2v_s}\phi^3, 
\end{multline}
\begin{multline}
\mL_4\approx
-\frac{m_h^2}{8v_{\rm ew}^2}h^4-\frac{m_h^2}{2v_{\rm ew}^2}\theta h^3\phi\\
-\frac{m_\phi^2-m_h^2}{4 v_{\rm ew}v_s}\theta h^2 \phi^2
+\frac{m_\phi^2}{2v_s^2}\theta h \phi^3-\frac{m_\phi^2}{8v_s^2}\phi^4, 
\end{multline}
in which $h$ and $\phi$ should be understood as physical particles.

\section{The complete expression of the $V_{\rm eff}(\phi,T')$}\label{app:VT}

The finite temperature potential of $\phi$ consists of zero-temperature tree level potential $V_0(\phi)$ as given in Eq.~(3) of the main text, the one-loop Coleman-Weinberg correction
\be
    V_1(\phi)=\sum_{i=\phi,A'}\frac{n_iM_i^4(\phi)}{64\pi^2}\left(\log\frac{M_i^2(\phi)}{\mu_R^2}-C_i\right),
\ee
where $\mu_R=10$ MeV is the renormalization scale,
\be
M_\phi^2(\phi)=-\frac{m_\phi^2}{2}+\frac{3m_\phi^2}{2v_s^2}\phi^2,\quad M_{A'}(\phi)=g_X\phi,
\ee
are the field-dependent masses, $n_{\phi,A'}=1$, 3 and $C_{\phi,A'}=3/2$, $5/6$, respectively. Note that the contribution from the Goldstone $\eta$ is neglected to avoid the IR divergence~\cite{Espinosa:2011ax}. The counter term $\delta V(\phi)\equiv\delta\mu_s^2\phi^2/2+\delta\lambda_s\phi^4/4$ is determined by the condition that one-loop zero temperature potential $V_0(\phi)+V_1(\phi)+\delta V(\phi)$ should still have a VEV at $\ave{\phi}=v_s$ with a mass eigenvalue $m_\phi$.

The thermal correction at one-loop level is given by
\begin{equation}
    V_{T1}(\phi,T')=\sum_{i=\phi,\eta,A'}\frac{n_iT'^4}{2\pi^2}J_B\left(\frac{M_i^2(\phi)}{T'^2}\right),
\end{equation}
where $n_\eta=1$ and $M_\eta^2(\phi)=-m_\phi^2(1-\phi^2/v_s^2)/2$, and the Bose thermal integral
\begin{equation*}
J_{B}(y)=\int_0^\infty x^2{\rm d} x\log(1- e^{-\sqrt{x^2+y}}).    
\end{equation*}
The thermal correction is dominated by the $A'$ contribution, as $g_X^2\gg\lambda_s$ in the parameter space of interest. The last term, i.e. the daisy resummation, is
\begin{equation}
  V_{\rm daisy}(\phi,T') = -\frac{g_X^3 T'}{12\pi} \left((\phi^2 + T'^2)^{3/2} - \phi^3\right),
\end{equation}
where the sub-leading contributions from $\phi$ and $\eta$ are dropped.

\bibliographystyle{apsrev}
\bibliography{reference}

\end{document}